# "DESIGN AND DEVELOPMENT OF DESKTOP BRAILLE PRINTER AT FABLAB NEPAL"

[a]Daya Bandhu Ghimire, [b]Pallab Shrestha,

[ab]Fablab Nepal

[a]075bme016.daya@pcampus.edu.np

[b]pallab.shrestha@impactub.net

The development of a desktop Braille printing machine aims to create an affordable, user-friendly device for visually impaired users. This document outlines the entire process, from research and requirement analysis to distribution and support, leveraging the content and guidelines from the GitHub repository.

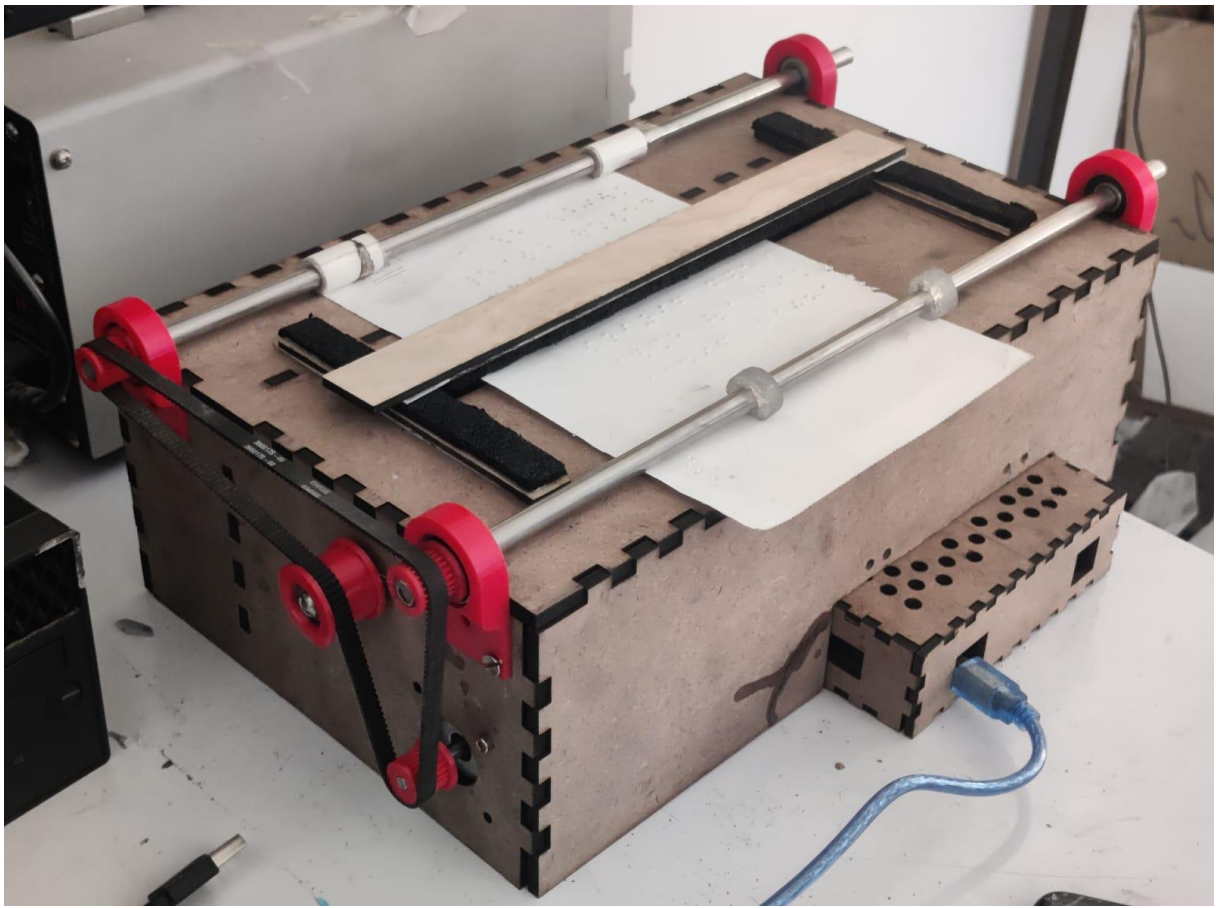

1.  **INTRODUCTION**

Of the approximately 36 million blind individuals worldwide, just over 7 million are under the age of 50 (2024, perkins.org). Braille is a system of raised dots designed to enable visually impaired individuals to read and write. Although commercial Braille printers are available, they are prohibitively expensive. Consequently, there is a pressing need for more affordable alternatives to these commercial Braille printers.

Braille script consists of a set of dots that correspond to specific letters, forming a unique language. Although there are various types of Braille scripts, we have identified a 3x2 matrix of 6 dots, as illustrated in Figure 1 below.

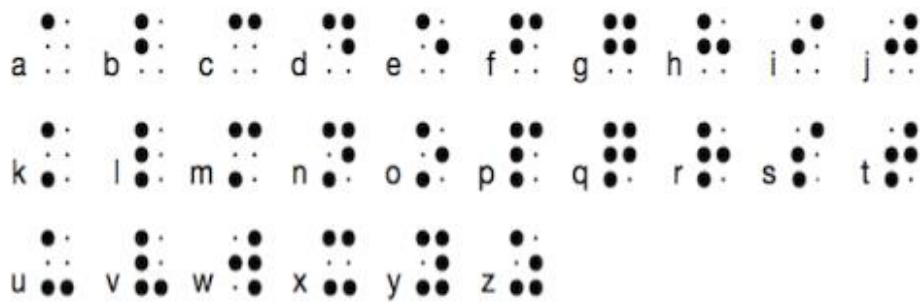

Figure 1 Braille Script

The six-dot Braille Unicode system is implemented with standard dimensions, allowing users to print up to sixty-four unique codes. The six-dot Unicode pattern and its dimensions are depicted in Figure 2 below.

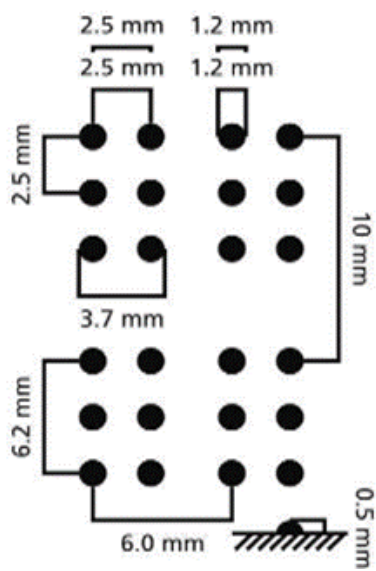

Figure 2 Six dot pattern

This project aims to repurpose spare parts from old printers to develop a functional desktop Braille printing machine. The machine is designed to print Braille script on paper by implementing CNC features. Its primary mechanism involves embossing dots onto the paper using a pointed pin-type actuator. The Braille printing machine employs the tinyG (ATMLE xmega192a3u) hardware as its main controller, ensuring precise control and functionality. The machine operates with two stepper motors that manage the X and Y axes, providing accurate positioning during the printing process. It accepts G-code via USB ports and interprets it for operation. A solenoid linear actuator is used to emboss Braille dots onto the paper. The use of ChiliPeppr software allows for real-time visualization of the work in progress, enhancing user interaction and monitoring capabilities.

## 2. METHODS

This section presents the framework for development of desktop braille printing machine. As illustrated in Figure 3 below. The methodology starts with requirement analysis and moves through prototyping to model final design after tests and refinements.

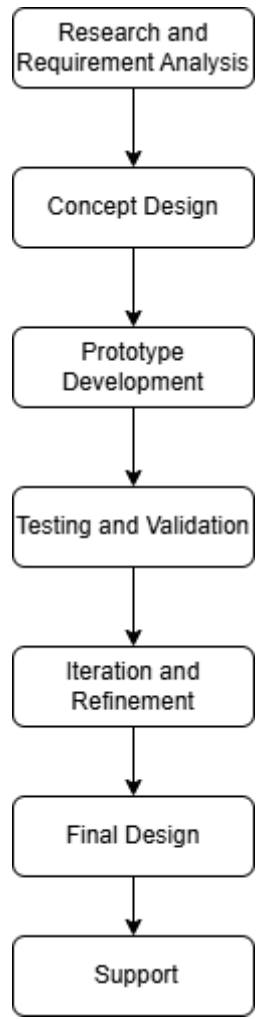

Figure 3: Schematic Overview

### 2.1. Research and Requirement Analysis

#### 2.1.1. User Needs Assessment

Understanding user needs is critical for developing a successful Braille printer. Through interviews and surveys with visually impaired individuals and organizations; and through internet research, key requirements were identified:

- Portability: Users need a lightweight, compact device for easy transportation.
- Ease of Use: The machine should be simple to operate, with intuitive controls.
- Affordability: Cost is a significant factor, as many existing Braille printers are prohibitively expensive.

#### 2.1.2. Market Analysis

A comprehensive market analysis was performed to understand the current landscape of Braille printers, identifying limitations such as high cost and complexity. This analysis highlighted areas for improvement, guiding the development of a more accessible and user-friendly device.

## 2.2. Concept Design

### 2.2.1. Brainstorming Sessions

Collaborative brainstorming sessions were held to generate innovative ideas for the Braille printer with the help of user groups and fellow engineers. Ideas focused on:

- Innovative Mechanisms: Exploring different mechanisms for creating Braille dots.
- User Interface: Designing an intuitive interface for easy operation.

### 2.2.2. Initial Sketches and CAD Models

Initial sketches and CAD models were created to visualize potential designs. These models helped in evaluating the feasibility of different concepts and provided a foundation for detailed design work. Key considerations included:

- Ergonomics: Ensuring the design is comfortable and easy to use.
- Space Utilization: Efficiently using space to keep the machine compact.
- Component Placement: Strategically placing components for optimal performance and maintenance.

## 2.3. Prototype Development

The initial prototype involved using parts from old 3D printers. Simple modifications such as attachment of solenoid unit was done along with integration of electronics and bed. Mountings and wirings are connected for the first prototype.

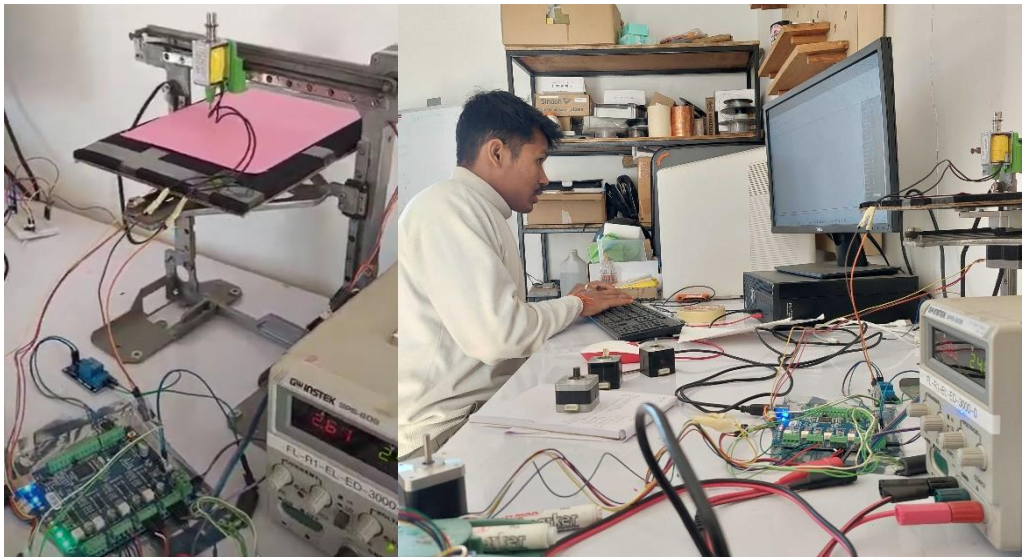

Figure 4: Prototype development

After all structural components and supporting attachments are joined. The initial power up and functional testing of each component are done. Integrating the electronic components was the next step, which involved mounting sensors and joining actuators.

## 2.4. Testing and Validation

### 2.4.1. System Testing

System testing was done focusing on ensuring the structural integrity and reliability of the prototype. Load application and forces are applied to test durability of bearing, embossing pins and 3D printed parts.

Motors are tested before wiring. It is advisable to never connect or disconnect anything (except possibly USB) with the power on. After testing motors, all motors are tested in complete configuration with a solenoid actuator, limit switch and relay switch. All the components are configured/controlled individually using the tinyG controller.

Motor->done

Solenoid->Done

### 2.4.2. Working with Interfaces

#### 2.4.2.1 Coolterm

Coolterm is a simple type interface to connect to the tinyG board. Once CoolTerm is set up, it acts as a terminal emulator, providing command-line access to TinyG and enabling file streaming to TinyG. The figure below illustrate the connection setup after the installation of CoolTerm is completed

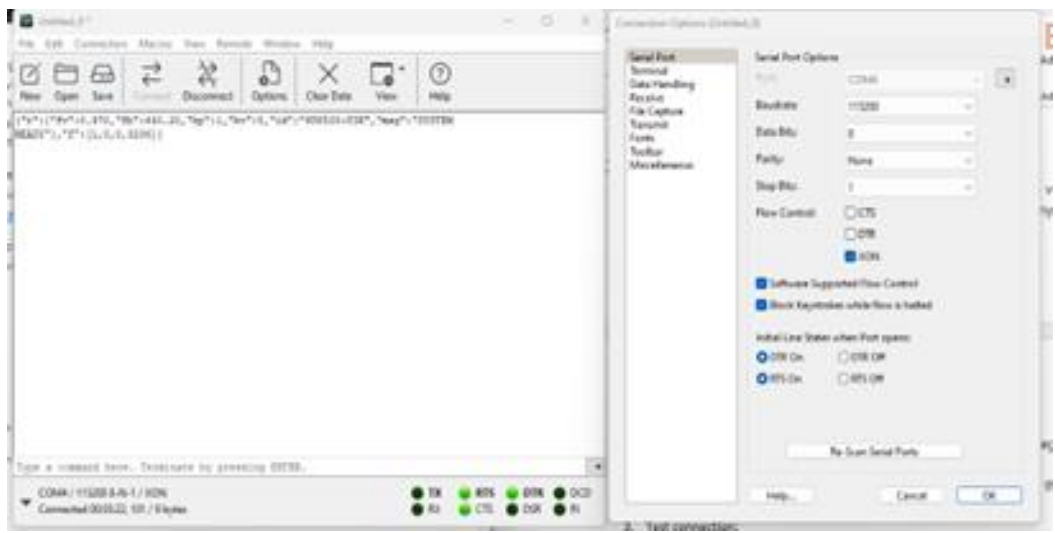

Figure 5: Coolterm Interface

*Since the control and Gui preview to g code is not present in coolterm, a new interface is used.

#### 2.4.2.2 Chillipeppr

Chilipeppr is used for connecting the tinyG board. ChiliPeppr provides a range of essential features for controlling CNC machines and other hardware interfaces. These include a 3D Viewer for visualizing designs and operations, a Gcode Sender for transmitting instructions, and a TinyG internal stats viewer that monitors the planner buffer and other performance metrics. Additionally, ChiliPeppr offers flow control capabilities via the planner buffer and supports a Serial Port JSON server that is compatible with various operating systems including Windows, Mac, Linux, and Raspberry Pi, enhancing its versatility and usability across different platforms.

Before using chillipepper, the following tasks should be done:

- Download Jason port server
- Run an interface of the port connected to this web app.
- Familiarize with chillipepper tinyG interface

The parameters that are needed for motor setup in chillipeppr are :

Figure 6: Motor Parameters

Also, axis parameters are needed:

Figure 7: Axis Parameters

This chillipeppr is very useful and we will continue using chillipeppr as a computer aided manufacturing software for the braille printer.

## 2.5. Iteration and Refinement

After the first prototype was ready for the, first braille was printed.

Test 1 :

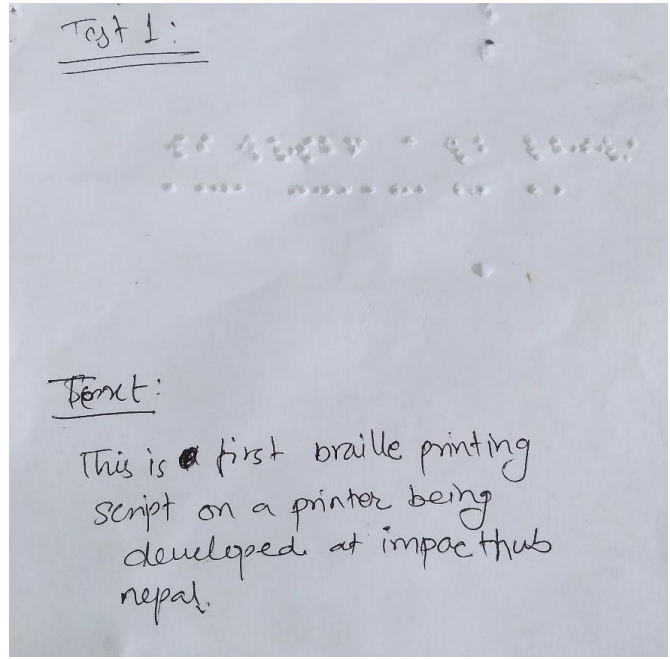

Figure 8: Test 1

| Problem Observed | Rectification |
| --- | --- |
| Correct Dimension of braille is needed. | New size and dots spacing are opted |
| Motor Vibration | Stepper motor parameter is adjusted. |

Test2:

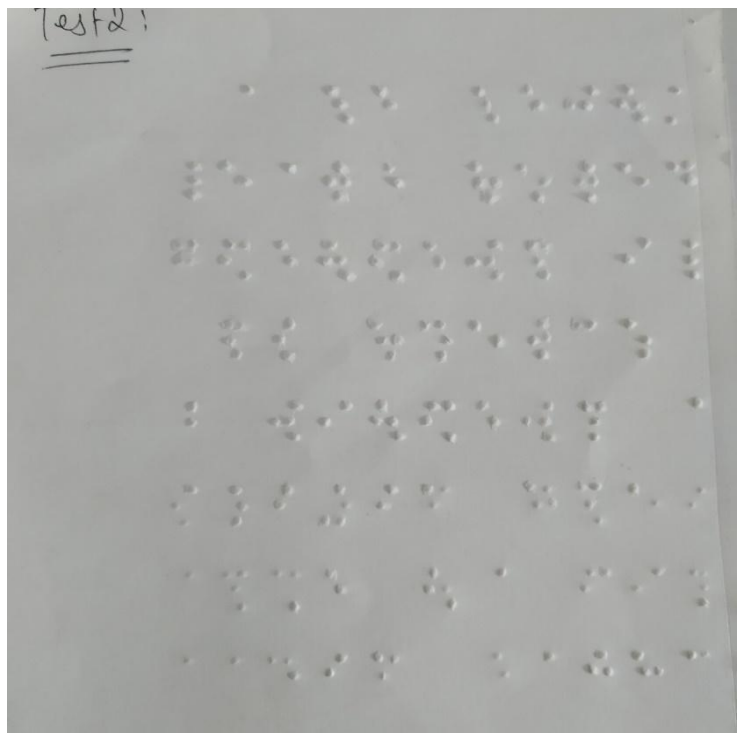

Figure 9: Test 2

| Problem Observed | Rectification |
|---|---|
| Non uniform Embossing | Modify the bed orientation |
| Motor Heating | Current rating is adjusted |

Test3:

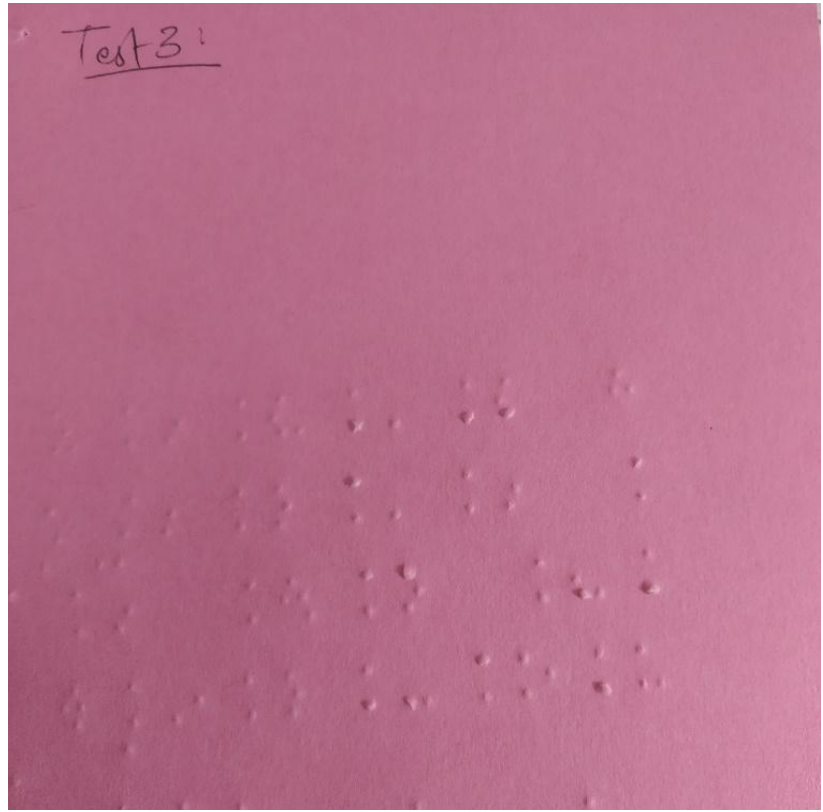

Figure 10: Test 3

| Problem Observed | Rectification |
|---|---|
| Less penetration | Proper thickness of paper is used. |

Test 4:

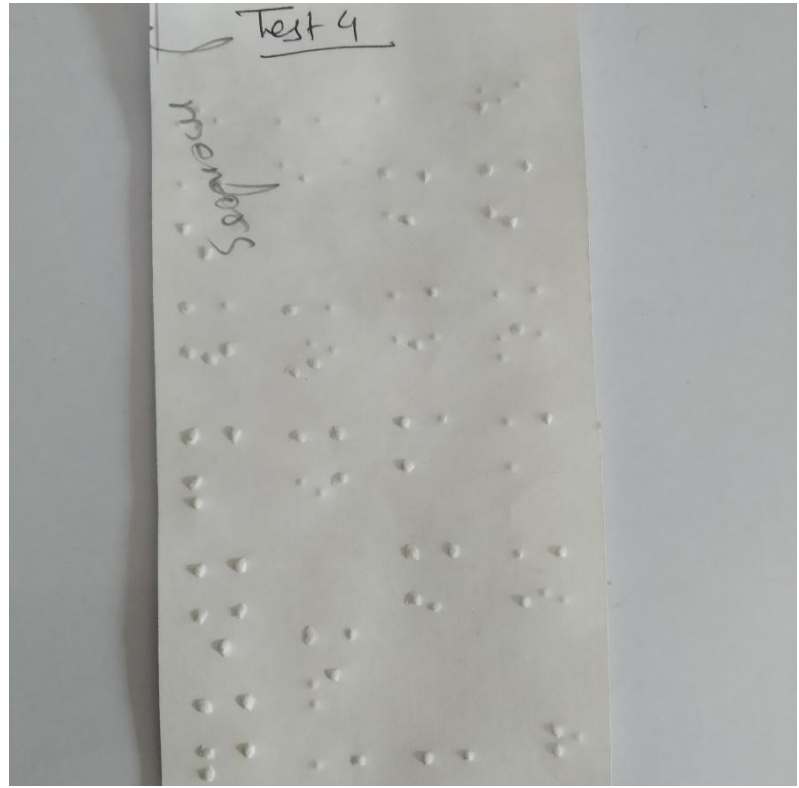

Figure 11: Test 4

| Problem Observed | Rectification |
|---|---|
| Vertically mirrored letter | The solenoid should engrave from upside down , so next version is developed. |

Hence, further design modifications are needed to do for the final design for production.

## 2.6. Final Design

### 2.6.1. Mechanical Design

The mechanical design phase involved creating detailed drawings and specifications for the printer's structural components. Key elements included:

#### 2.6.1.1 System Box and Housing

In this project, a system box and housing for a Braille printing machine were successfully designed and fabricated using laser cutting technology. The process began with conceptualizing and sketching the design while considering dimensions, component opening. A 3D model was then created using CAD software, serving as a digital blueprint for fabrication. Suitable materials were selected for each component, ensuring compatibility with the laser cutter. The design was optimized for laser cutting, with tolerances, material thickness, and kerf considered. Individual components were prepared as 2D vector files and imported into the laser cutter's software for fabrication. Each component was carefully cut, cleaned, and inspected for defects before assembly.

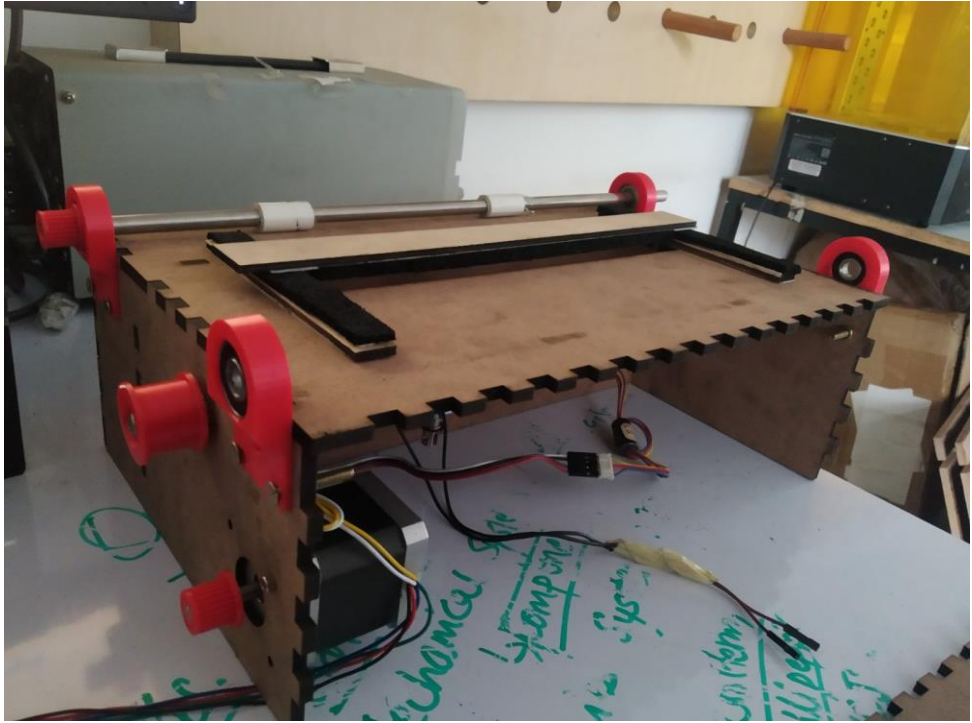

Figure 12: System Box and Housing

During assembly, components were slotted together in a sturdy structure. Finishing touches, such as sanding, painting, and labeling, were applied, followed by testing the Braille printing machine in its new enclosure, confirming proper functioning and design requirements were met.

### 2.6.1.2 Shafts and Paper Rollers

A stainless-steel shaft and paper rollers were successfully fabricated using cutting and fitting techniques for the shaft and silicon molding for the rollers. Th e process began with sourcing a stainless-steel rod for the shaft. The rod was precisely measured and marked according to the required dimensions. Using a metal cutting saw, the rod was carefully cut to the desired length. Afterward, a bench grinder was employed to smooth the edges and ensure a proper fit with the roller components. For the paper rollers, a silicon molding process was used. First, a master pattern was created using 3D printer with precise dimensions. This pattern was then used to create a silicone mold by pouring liquid silicone rubber over the pattern and allowing it to cure.

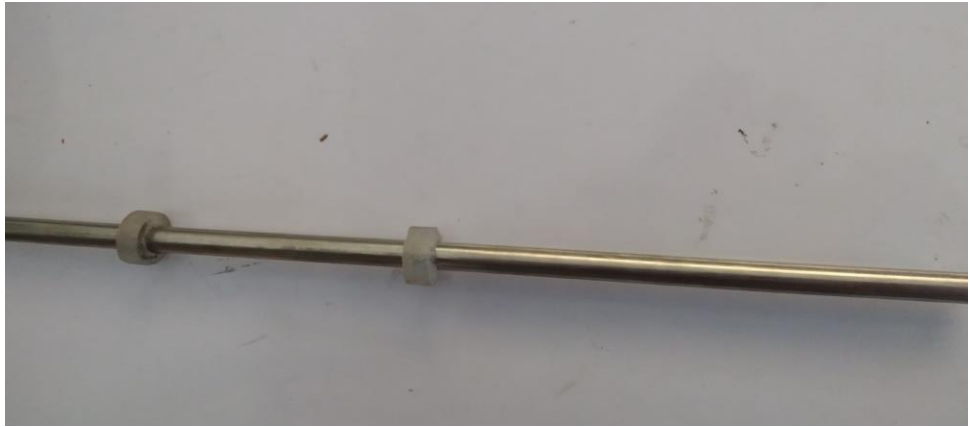
Figure 13: System Box and Housing

### 2.6.1.3 Bearing and Housing

A bearing and bearing housing were successfully integrated using a purchased bearing and a 3D printing technique for the bearing housing. The process began with selecting an appropriate bearing based on the required specifications for the project. Once the bearing was acquired, the focus shifted to designing and fabricating a suitable housing. A 3D model of the bearing housing was created using CAD software, ensuring that the design would securely hold the bearing in place while allowing for smooth rotation. Once the 3D model was finalized, it was exported as a printable file and sent to a 3D printer. Using a suitable filament material, such as PLA, the bearing housing was printed layer by layer until the complete part was formed. After printing, the housing was carefully removed from the build plate, and any support structures were cleaned off. Next, the bearing was test-fitted into the housing to ensure a proper fit.

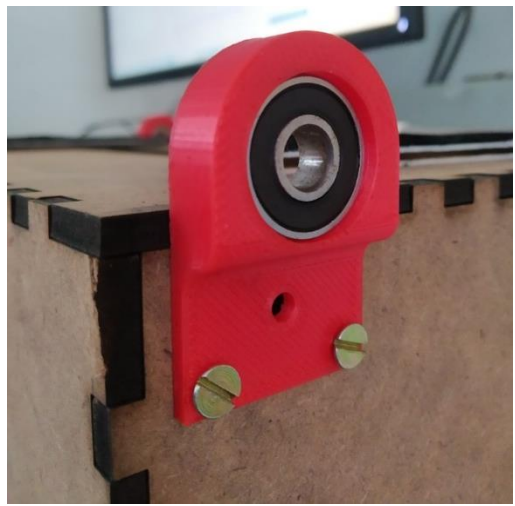
Figure 14: Bearing and Housing

### 2.6.1.4 Pulley and Belts

Next, a suitable belt was selected and purchased. Then, A 3D model of the pulley was created using CAD software, ensuring that the design matched the specifications of the belt, such as pitch, width, and diameter. Once the 3D model was finalized, it was exported as a printable file and sent to a 3D printer. Using a suitable filament material, such as PLA, the pulley was printed layer by layer until the complete part was formed. After printing, the pulley was carefully removed from the build plate, and any support structures were cleaned off. The belt was then test-fitted onto the 3D-printed pulley to ensure a proper fit and smooth rotation.

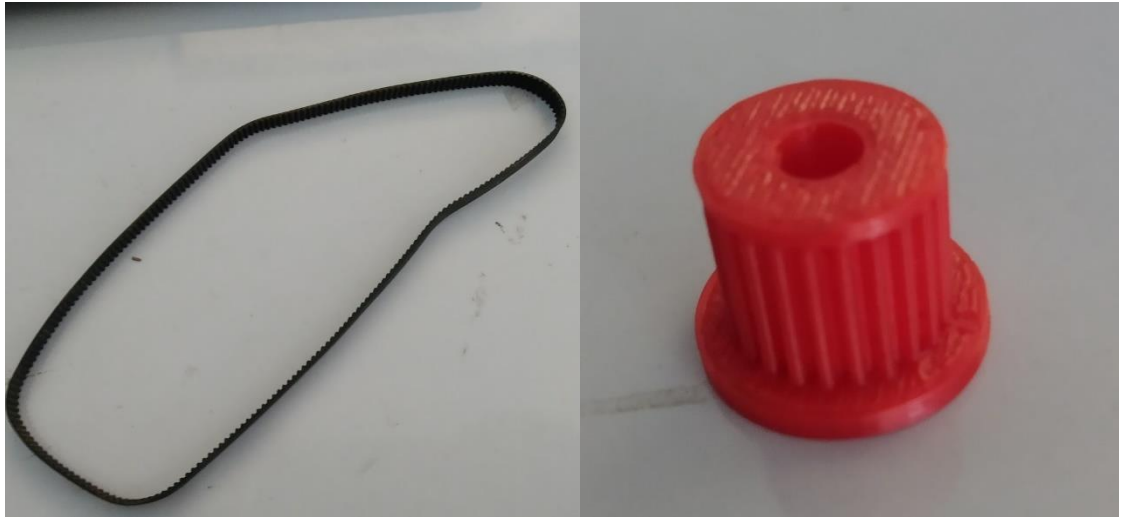

Figure 15 : Pulleys and belt

### 2.6.1.5 Rails and Pulleys
Rails and pulleys from old 3D printers are reused.

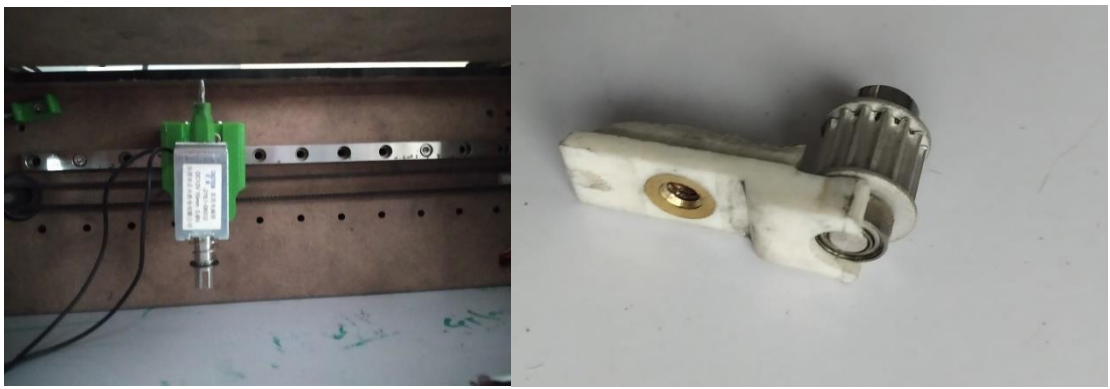

Figure 16: Rails and Pulleys

### 2.6.1.6 Linear bearing and Housing
Linear bearing from old 3D printer is used and the housing is modeled in CAD and then 3D printed.

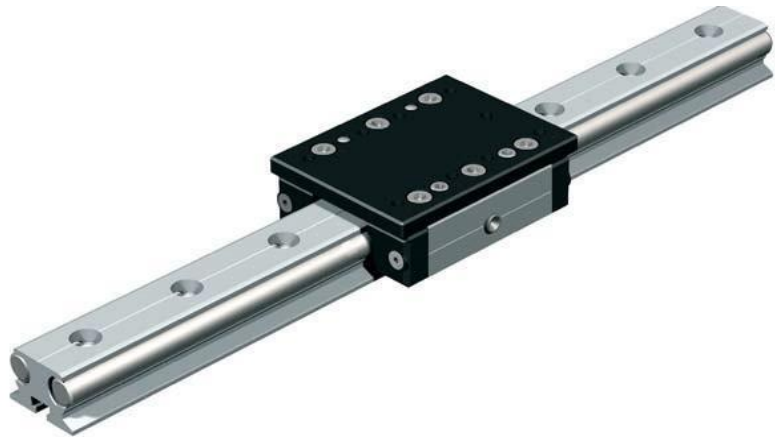

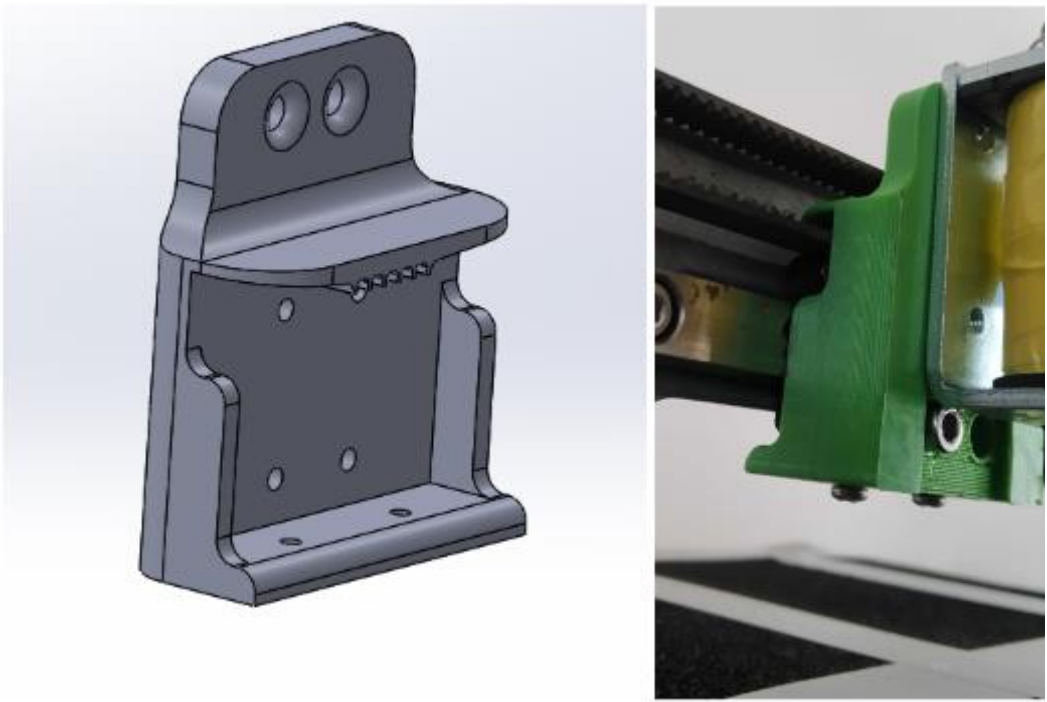

Figure 17: Linear Bearing and Housing

### 2.6.1.7 Limit switch handler
The limit switch coupler with the machine is designed and 3D printed.

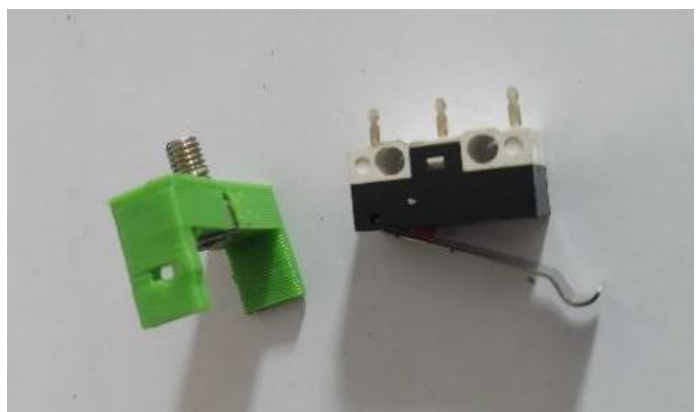

Figure 18: Limit Switch Handle V1

The first iteration didn't give a good outcome as shrinkage of the hole was not considered. Also, the line thickness was 0.3mm hence it produced a bad quality print. It was structurally weak also.

The second iteration was done , it produced a structurally sound print considering shrinkage effect.

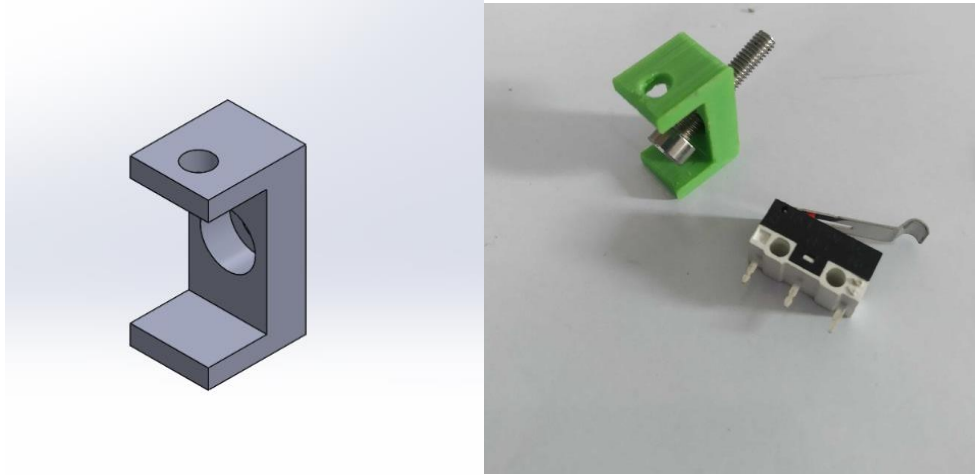

Figure 19: Limit Switch Handle V2

### 2.6.2. Electronics Systems

The basic layout of electronics circuit is presented below.

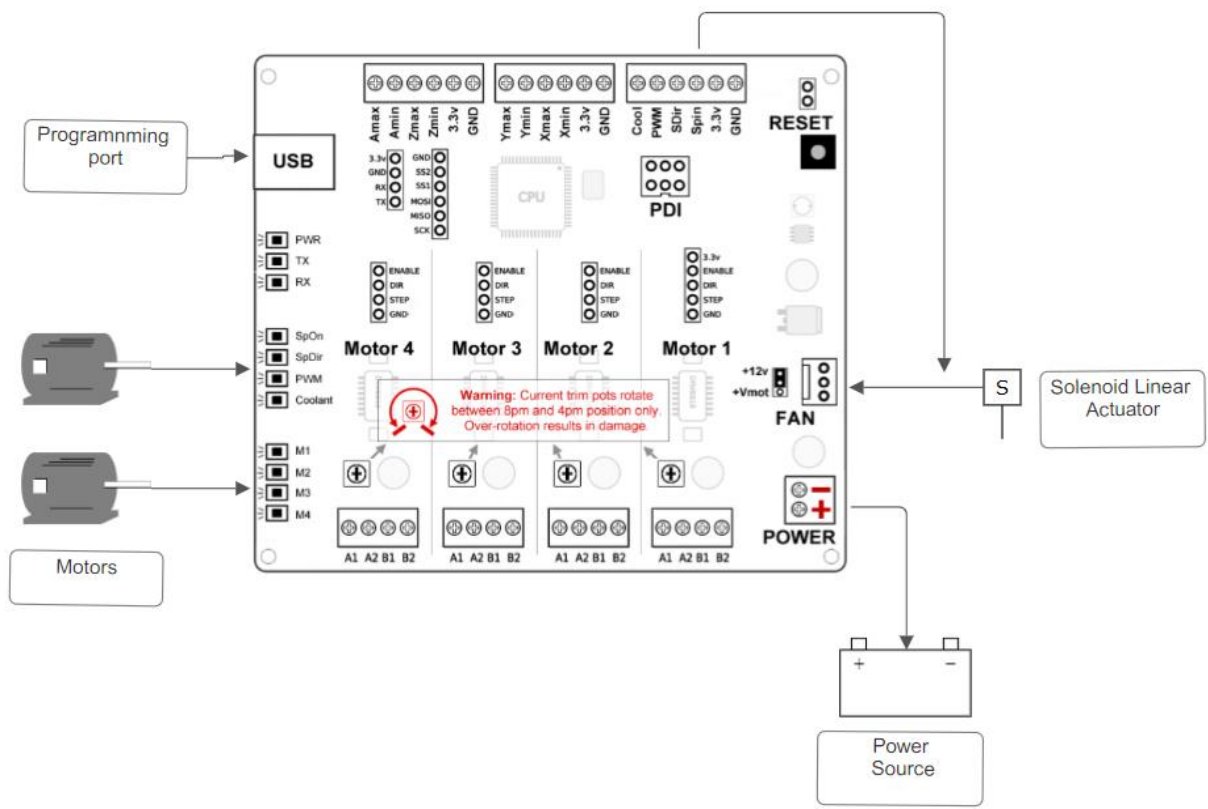

Figure 20: Electronic System Layout

The main controller is TinyG along with other accessories. The basic electronic components are:

#### 2.6.2.1 TinyG v8 board

The TinyG project is a multi-axis motion control system. It is designed for CNC applications and other applications that require highly precise motion control. TinyG is meant to be a complete embedded solution for small/medium motor control.

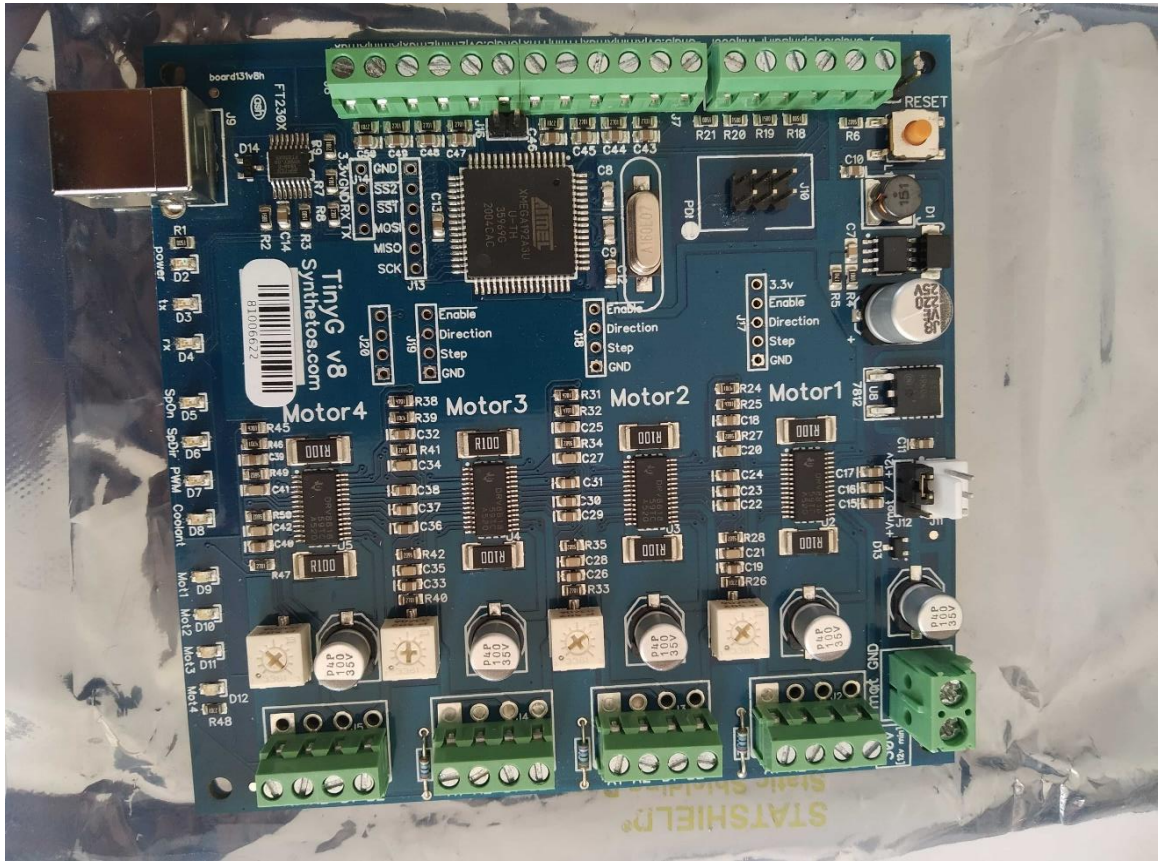

Figure 21: TinyG v8 board

While operating the board, different difficulties are encountered and solved simultaneously. One the problem encountered was :Spdir was continiously flashing error.This problem was solved using unlocking bits using ATmel-ice programmer as stated in the link.

#### 2.6.2.2 Controller box

For storing the tinyG board, a controller box is developed.

Version 1: 3D printed controller box is developed.

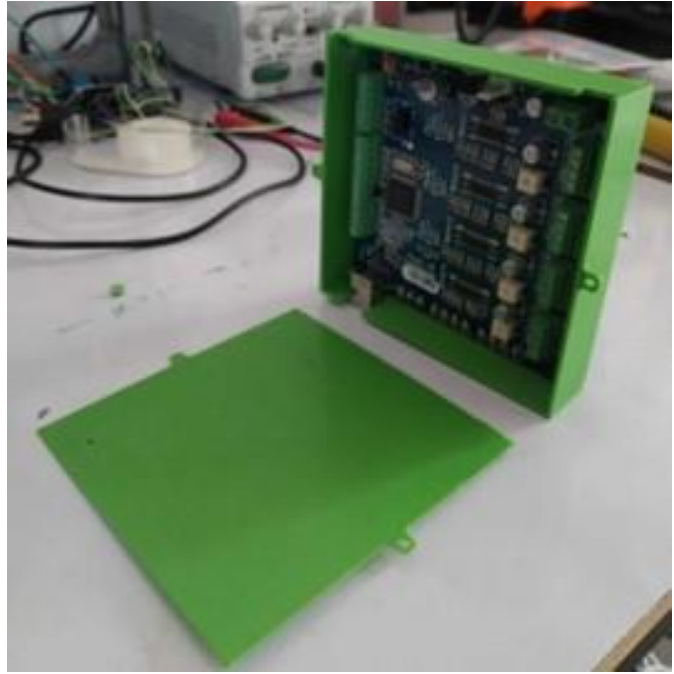

Figure 22: Controller box V1

Version 2: Laser cut box is used to house the board and limit switch also.

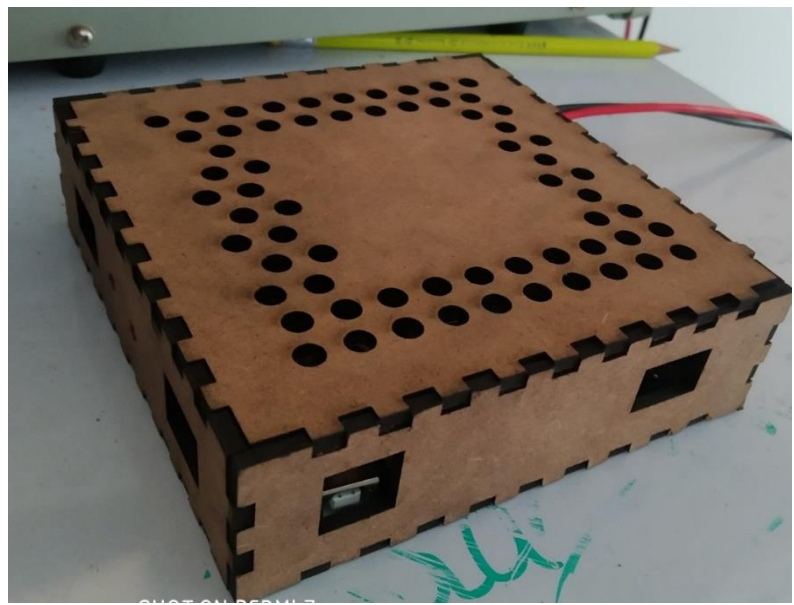

Figure 23: Controller box V2

### 2.6.2.3 Nema17 motors and parameters

The stepper motors that we have used are 17PM-K374BN01CN model, bipolar type joined with the driver of TinyG board. The specifications of motors are shown below:

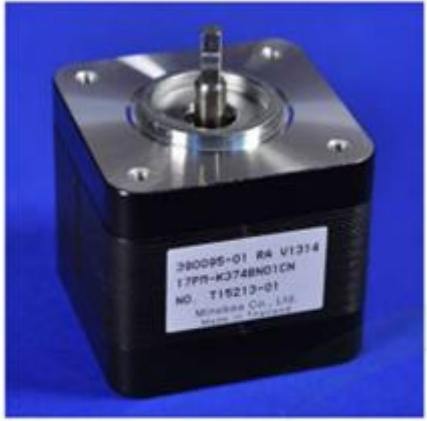
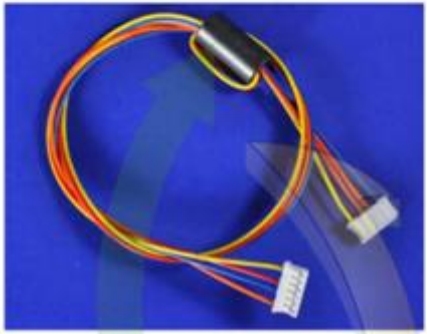
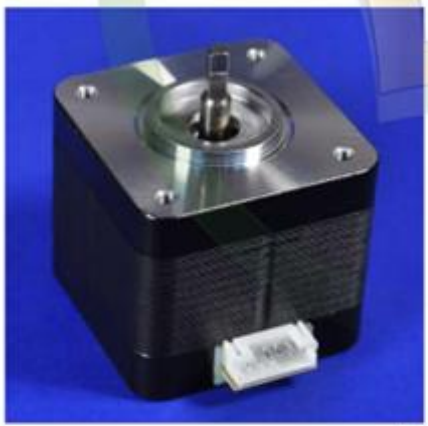
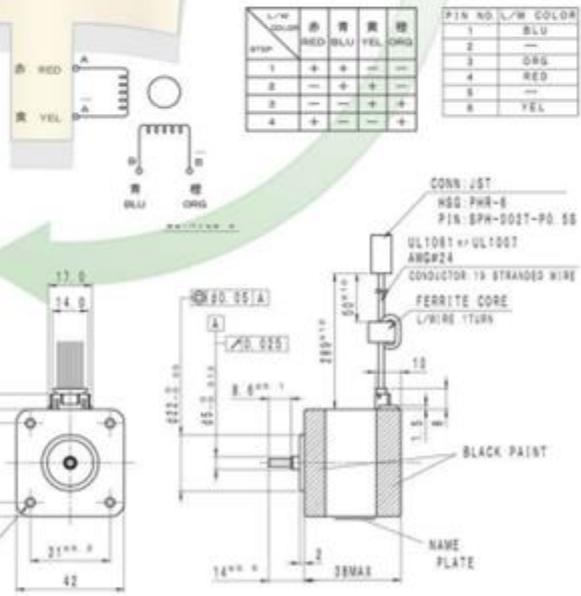

Figure 24: Stepper Motor

### 2.6.2.4 Solenoid Linear actuator

For embossing the dots for braille script, solenoid actuator is used. 12V is needed to activate the solenoid which is controlled by signal.

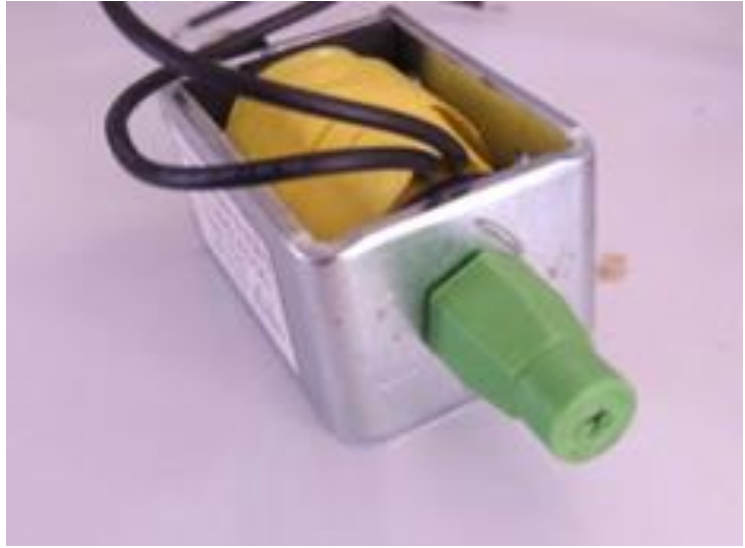

Figure 25: Solenoid Linear Actuator

### 2.6.2.5 Relay switch

Fan output from the TinyG is constant voltage output and has no control using the board. Hence, relay switch is used to provide voltage to solenoid and control the signal also.

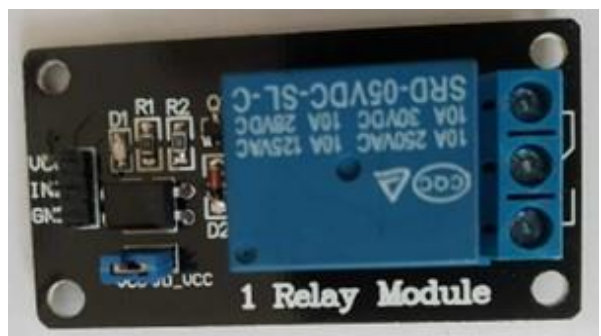

Figure 26: Relay Switch

### 2.6.2.6 Limit switch

For homing the motors, limit switch is used at the end of axis terminal. Limit switch definition is an electromechanical switch that operates by any physical force or the movement of a machine. These switches are very helpful in detecting the absence or presence of an object, counting, detecting speed, detecting movement range, travel limit, positioning, etc. These switches include three terminals NO (Normally Open), NC (Normally Open) & Common.

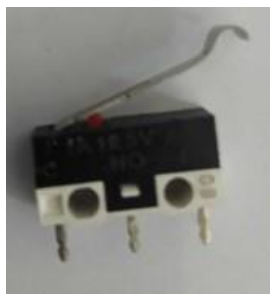

Figure 27: Limit Switch

After all the system are identified and developed, final Integration is done and braille is printed.

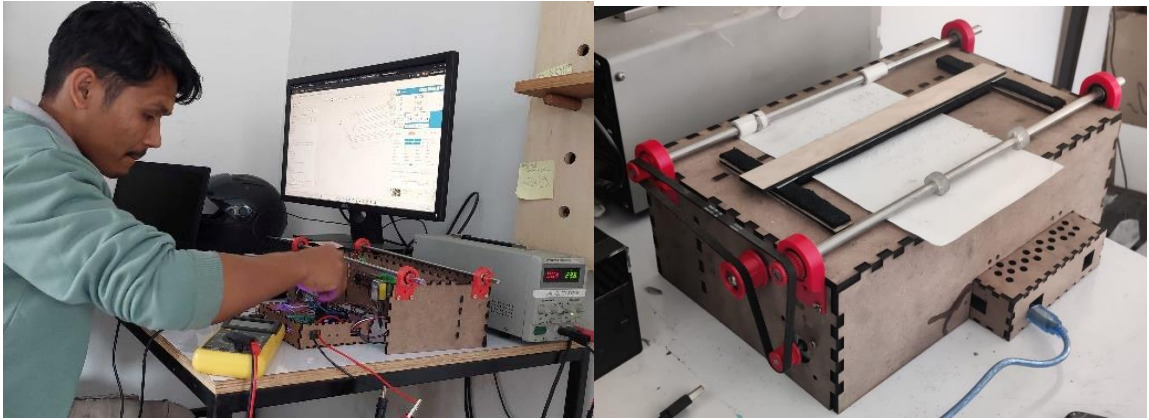

Figure 28: Complete Assembly

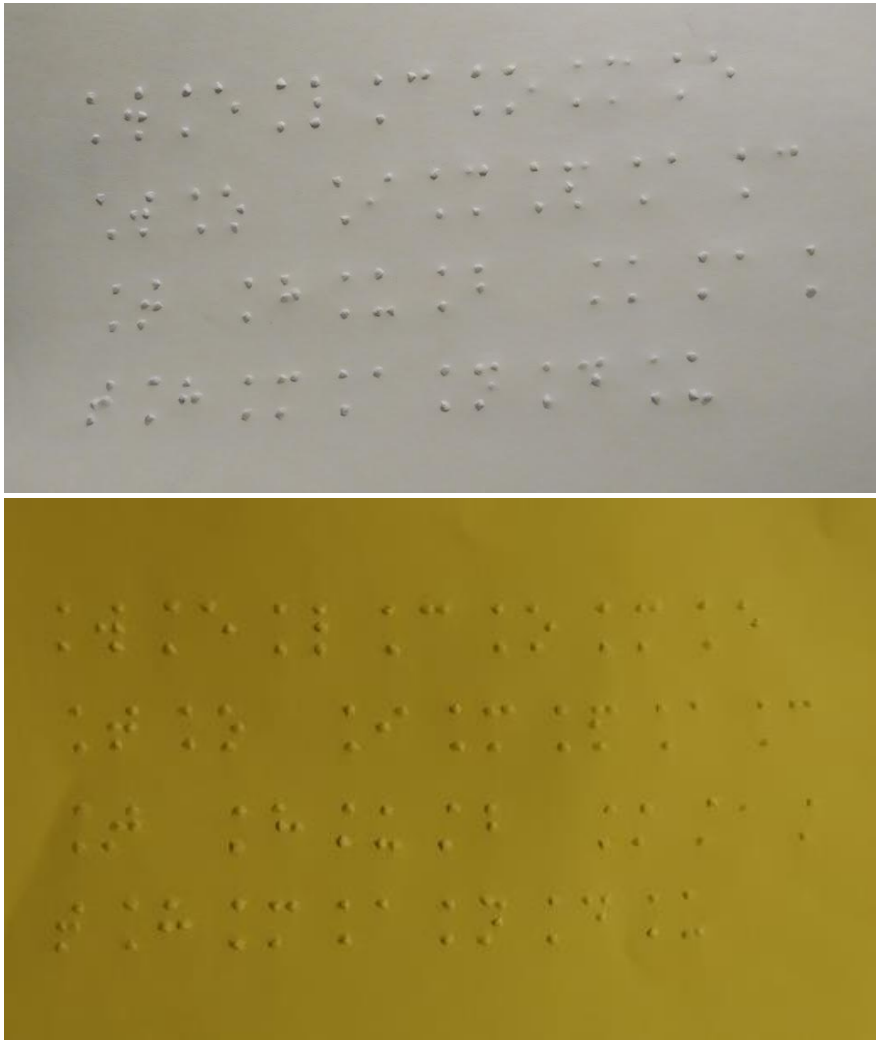

Figure 29: Final Braille Prints

## 2.7. Support

After the machine is assembled, following operational steps must be followed to print a braille script on a paper.

### 2.7.1. Operational Steps

After the machine has been assembled, the following steps are done to print on braille script:

- ➢ Use Braillerap to generate Gcode .

This site converts English to Braille and generates g code on the basis of a given text. We can define the size of the paper on this site also.

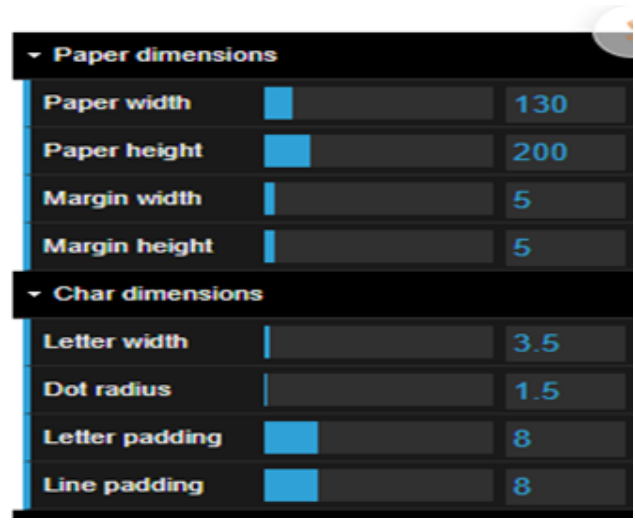

Figure 30: Paper and Character Dimension

- ➢ Conditioning and reviewing Gcode for correctness.

- ➢ Power the braille printer and connect it to the interface.

A DC power supply between 12 and 30 volts, 24 volts is ideal. It should be capable of providing 4 to 15 amps. Before connecting it to TinyG, turn on the power supply and make sure you have the correct voltage. Incorrect voltage or polarity can destroy the board!

- ➢ Load Gcode into the interface and send it to the machine.

- ➢ Load the paper, manage it, and print on the paper.

## 3. CONCLUSION

To conclude, design, assembly and fabrication of braille printing prototype is done at fablab Nepal lab under Impacthub Kathmandu. This project has really helped to enhance the skills in product design, CAD AND 3D printing along with handling microcontrollers. It has given insight into CNC machining and the opportunity to build one. Overall, the project was successful.

## 4. FUTURE RECOMMENDATIONS

- This project has a huge potential for further work, design improvement, and innovation.
- The scale can be increased in terms of its work.
- There is a lot of space for design improvement and optimization. It can be redesigned to make it more cost-effective.

Files are available to download from : https://github.com/fablabnepal1/Desktop-Braille-Printing-Machine